\documentclass[a4paper,11pt]{article}
\usepackage{pos}
\usepackage{caption}
\usepackage{subcaption}

\title{AstroPix: CMOS pixels in space}

\author[a,b]{Amanda L. Steinhebel}
\author*[b]{Regina Caputo}
\author[c,b,d]{Henrike Fleischhack}
\author[e]{Nicolas Striebig}
\author[f]{Manoj Jadhav}
\author[g]{Yusuke Suda}
\author[f]{Ricardo Luz}
\author[a,b]{Daniel Violette}
\author[b]{Carolyn Kierans}
\author[h]{Hiroyasu Tajima}
\author[g]{Yasushi Fukazawa}
\author[e]{Richard Leys}
\author[e]{Ivan Peric}
\author[f]{Jessica Metcalfe}
\author[i,b,d]{Michela Negro}
\author[b]{Jeremy S. Perkins}

\affiliation[a]{NASA Postdoctoral Program Fellow}
\affiliation[b]{NASA Goddard Space Flight Center, Greenbelt, MD, USA}
\affiliation[c]{Catholic University of America, Washington, DC, USA}
\affiliation[d]{Center for Research and Exploration in Space Science and Technology, NASA/GSFC, Greenbelt, MD, USA}
\affiliation[e]{Karlsruhe Institute of Technology, Karlsruhe, Germany}
\affiliation[f]{Argonne National Laboratory, Lemont, IL, USA}
\affiliation[g]{Hiroshima University, Higashi-Hiroshima City, Hiroshima, Japan}
\affiliation[h]{Institute for Space-Environment Research, Nagoya University, Nagoya, Aichi, Japan}
\affiliation[i]{University of Maryland, Baltimore County, Baltimore, MD 21250, USA}

\emailAdd{amanda.l.steinhebel@nasa.gov}

\abstract{Space-based gamma-ray telescopes such as the Fermi Large Area Telescope have used single sided silicon strip detectors to measure the position of charged particles produced by incident gamma rays with high resolution. At energies in the Compton regime and below, two dimensional position information within a single detector is required. Double sided silicon strip detectors are one option; however, this technology is difficult to fabricate and large arrays are susceptible to noise. This work outlines the development and implementation of monolithic CMOS active pixel silicon sensors, AstroPix, for use in future gamma-ray telescopes. Based upon detectors designed using the HVCMOS process at the Karlsruhe Institute of Technology, AstroPix has the potential to maintain the high energy and angular resolution required of a medium-energy gamma-ray telescope while reducing noise with the dual detection-and-readout capabilities of a CMOS chip. The status of AstroPix development and testing as well as outlook for application in future telescopes is presented.}

\FullConference{%
  10th International Workshop on Semiconductor Pixel Detectors for Particles and Imaging (Pixel2022)\\
  12-16 December 2022\\
  Santa Fe, New Mexico, USA}

\graphicspath{{./}{figures/}}

\begin{document}
\maketitle

\section{Motivation and Previous Work}
\label{sec:intro}

\begin{figure}
    \centering
    \includegraphics[width=0.5\textwidth]{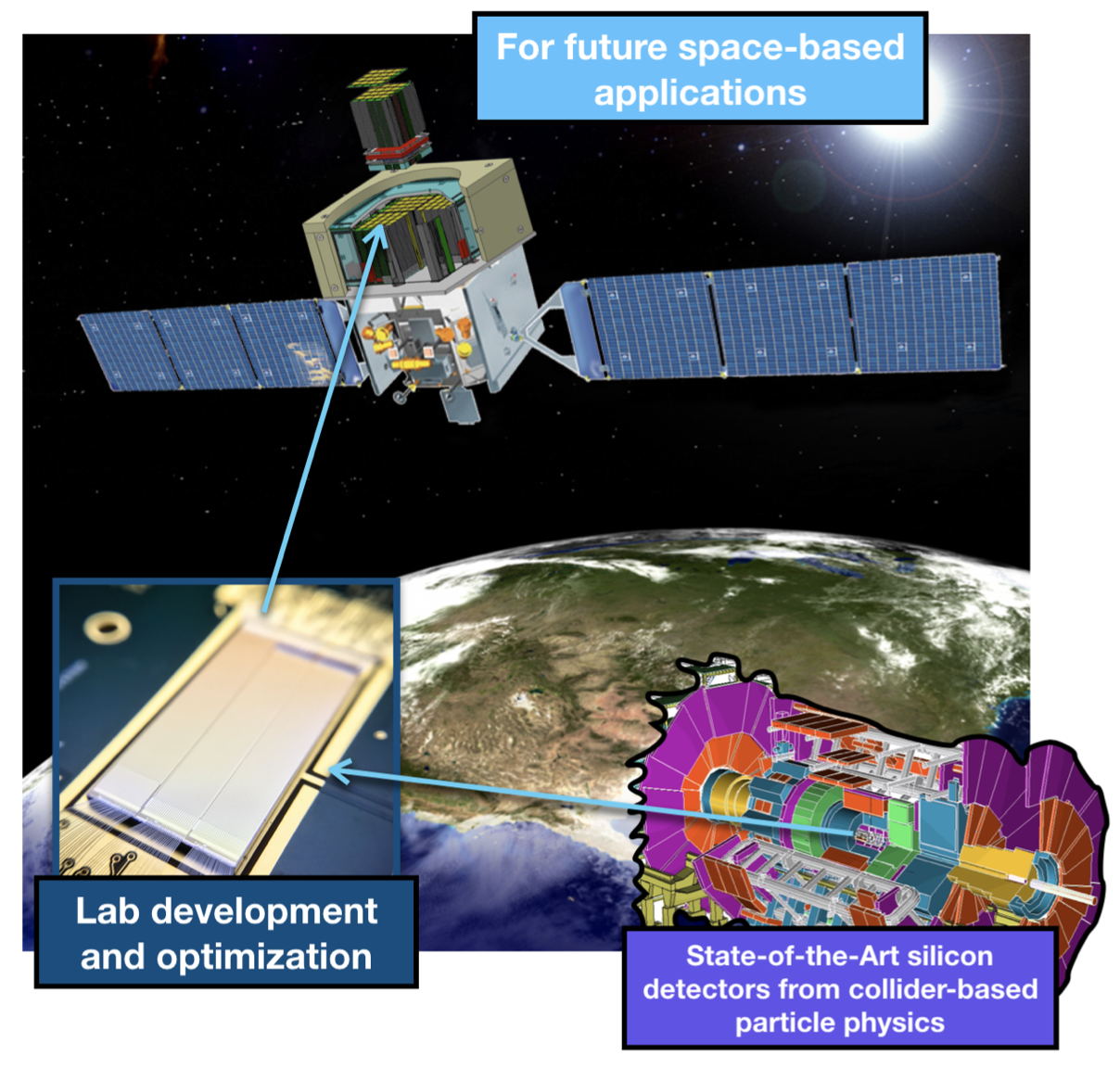}
    \caption{AstroPix project overview - collider-based particle physics silicon detectors have been redesigned for use in space and ongoing lab development and optimization will ready AstroPix for future space-based applications in gamma-ray tracker subsystems.}
    \label{fig:projectOverview}
\end{figure}

The AstroPix project aims to develop and test pixelated silicon sensors for use in space-based gamma-ray instruments. This novel space-based technology is based upon work done with similar detectors at the Large Hadron Collider \cite{schoning2020mupix}, and could contribute to a host of future instruments such as a next-generation wide-field gamma-ray explorer whose time domain capabilities is prioritized in the Astro2020 Decadal Survey \cite{decadal}. This work will overview the design, testing, and development of AstroPix version 2. This version is a step toward a flight prototype which will be realized with version 3 (Section~\ref{sec:future}). The long-term goal is to continue this development and testing in order to determine AstroPix performance and its suitability for use in space (Fig.~\ref{fig:projectOverview}).  

\begin{figure}
    \centering
    \includegraphics[width=0.5\textwidth]{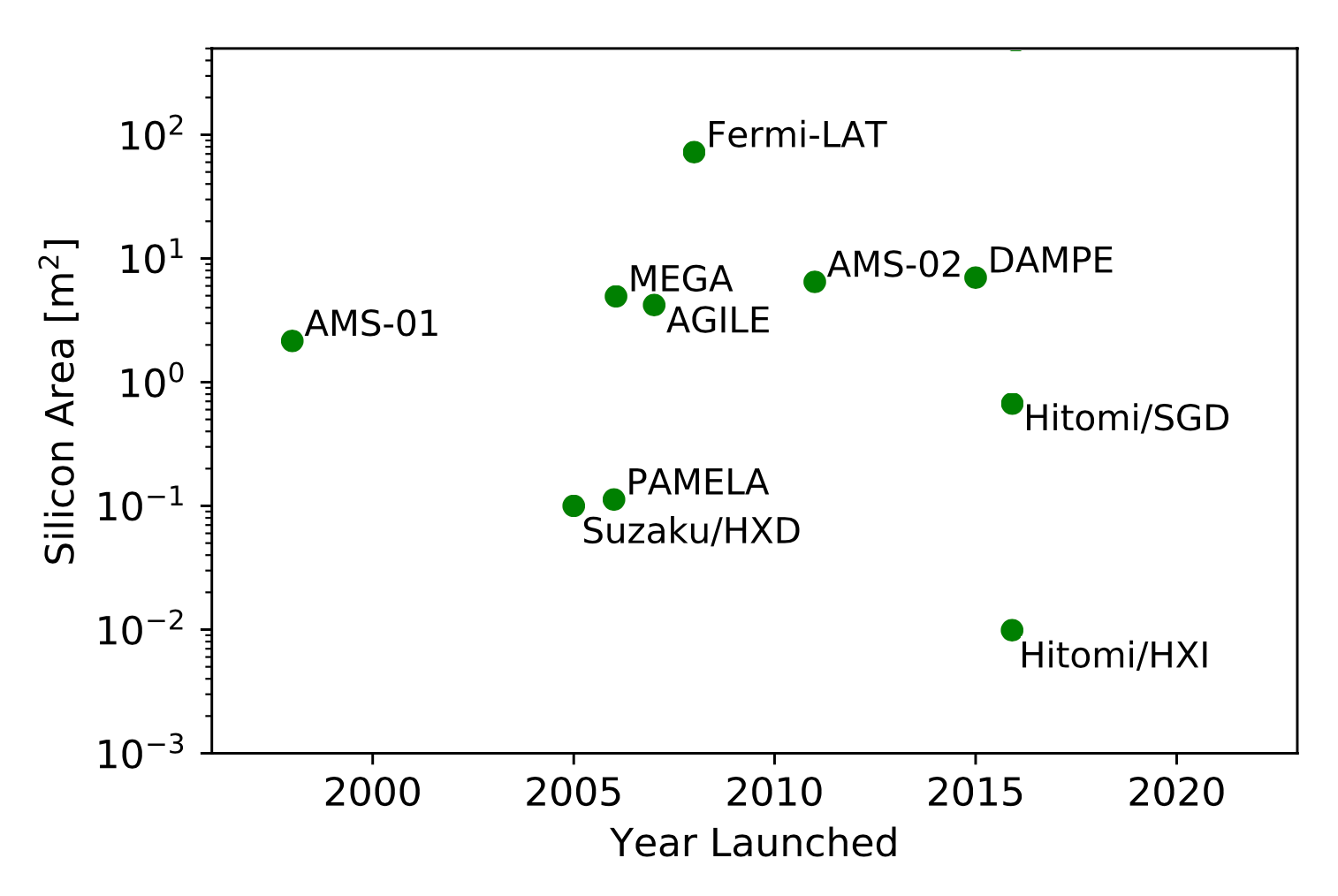}
    \caption{The varied use of heritage silicon detector technology in space and balloon astrophysical instruments, modified from~\cite{Caputo2022}.}
    \label{fig:missions}
\end{figure}

Heritage technology used for tracking instruments on previous gamma-ray, hard X-ray, and cosmic ray instruments include single-sided silicon strip detectors, double-sided silicon strip detectors (Fig.~\ref{fig:missions}), and other pixelated detectors. Each design carries unique strengths and weaknesses regarding event timing, position resolution, readout efficiency, noise, energy resolution, and power consumption. The evolution of silicon tracker technology used in for the study of astrophysics in space has been made possible in part through its long history of implementing technologies developed by ground-based particle physics experiments. 

\begin{figure}
\centering
\begin{subfigure}{0.47\textwidth}
  \centering
  \includegraphics[width=.8\linewidth]{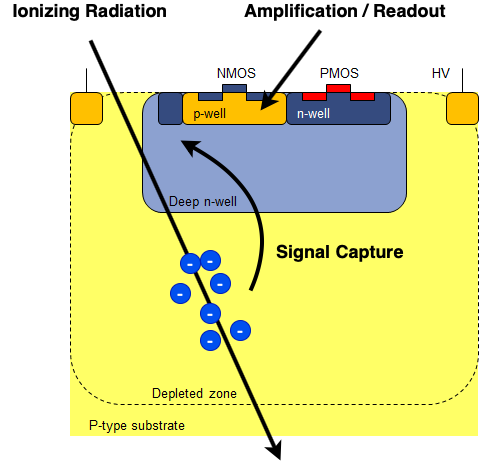}
  \caption{An illustration of how the HVCMOS pixel collects charge from an incoming particle. }
  \label{fig:cmos}
\end{subfigure}%
\qquad
\begin{subfigure}{0.47\textwidth}
  \centering
  \includegraphics[width=.8\linewidth]{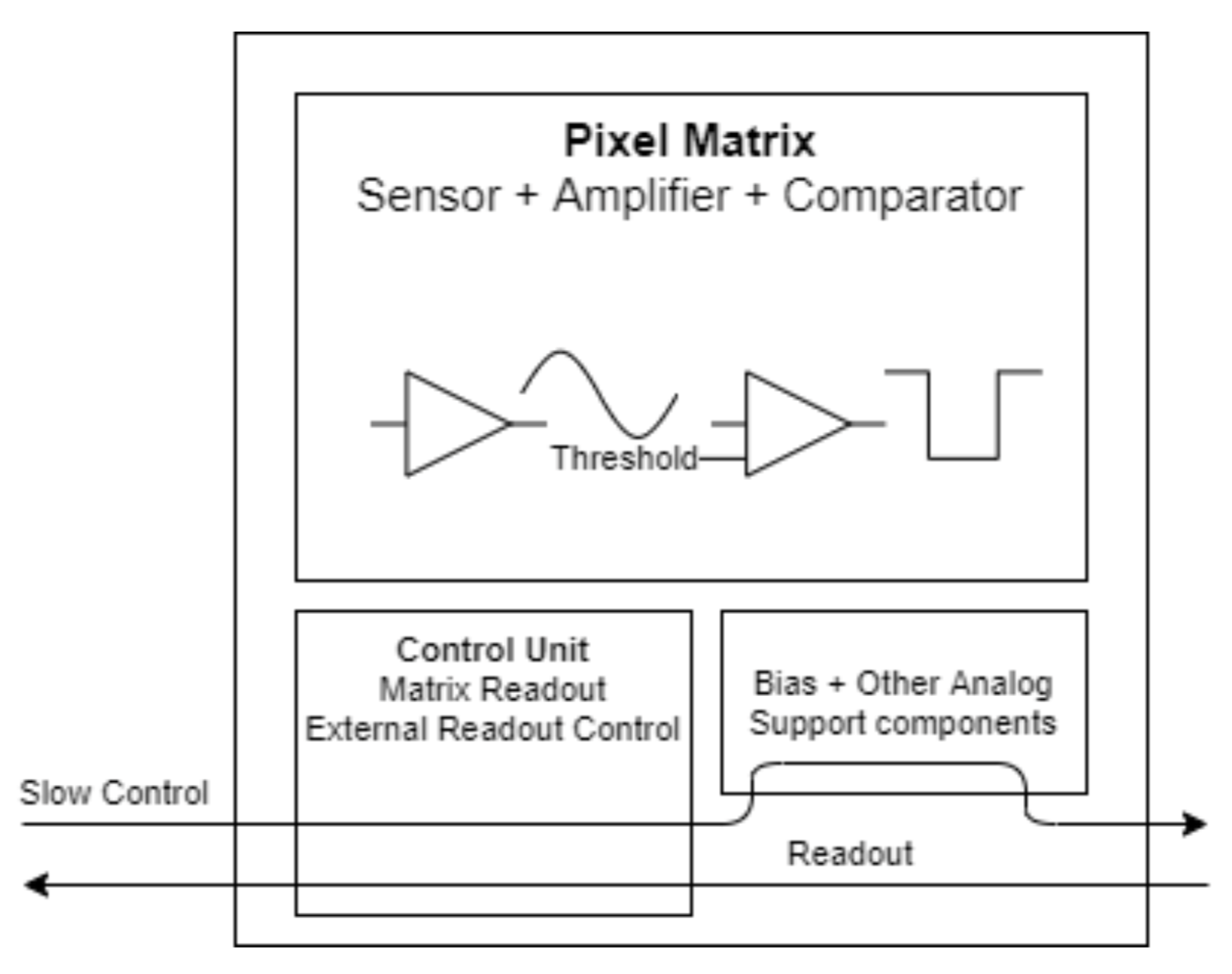}
  \caption{(Not to scale) The charge collection and amplification/comparator architecture are embedded in the pixel matrix. The matrix reads out to a control unit on the digital periphery for digitization and chip readout.}
  \label{fig:circuit}
\end{subfigure}%
\caption{HVCMOS design, where circuitry in each pixel allows for charge collection, amplification, and readout to the digital periphery of the chip \cite{striebig,Peric:2018lya}.}
\label{fig:eediags}
\end{figure}

The next step in the development of a silicon detector for space is AstroPix, a monolithic high voltage complimentary metal-oxide semiconductor (HVCMOS) sensor. CMOS pixels perform charge collection, signal amplification, and readout with electronics all co-integrated into the pixel matrix (Fig.~\ref{fig:eediags}). The addition of a high voltage bias to every pixel enhances the charge collection efficiency over previous diffusion-based methods (Fig.~\ref{fig:cmos}). At an individual pixel level, once the charge is collected, it is converted to a voltage signal by a charge sensitive amplifier, which goes into a comparator to generate a trigger above the threshold level. This signal is routed to the digital periphery of the chip (Fig.~\ref{fig:circuit}) where the output from all pixels are digitized and read out \cite{Peric:2018lya}. In this way, two signals can be tested - the analog signal from individual pixels being the output of the charge-sensitive amplifier, and the fully digitized full-chip digital signal as readout from the digital periphery. This analog data is used in testing and characterization, but final AstroPix designs will exclusively utilize digital data readout. 

An HVCMOS design such as AstroPix carries huge potential benefits over legacy technology in currently flying instruments. The CMOS fabrication process is common in commercial industry and well understood, making chip production affordable. Silicon is abundant, affordable, and operates at room temperature which further drives down costs. The CMOS design requires no readout ASIC board since the readout is done on-chip, easing integration and minimizing noise especially when compared to arrays of silicon strip detectors. The on-pixel circuitry is also customizable and low-power, creating sensors with less power per channel count relative to other pixelated or strip sensors.  

An HVCMOS chip designed for use at the Large Hadron Collider's ATLAS detector was first tested as a proof-of-concept study for AstroPix. This chip, called ATLASPix, utilized $150\times 50$~\um{}$^2$ pixels in four arrays of $25\times100$ pixels. Local testing detailed in \cite{Brewer:2021mbe} showed that ATLASPix was a feasible starting point for AstroPix development.  

\begin{figure}
\centering
\begin{subfigure}{0.47\textwidth}
  \centering
  \includegraphics[width=0.9\linewidth]{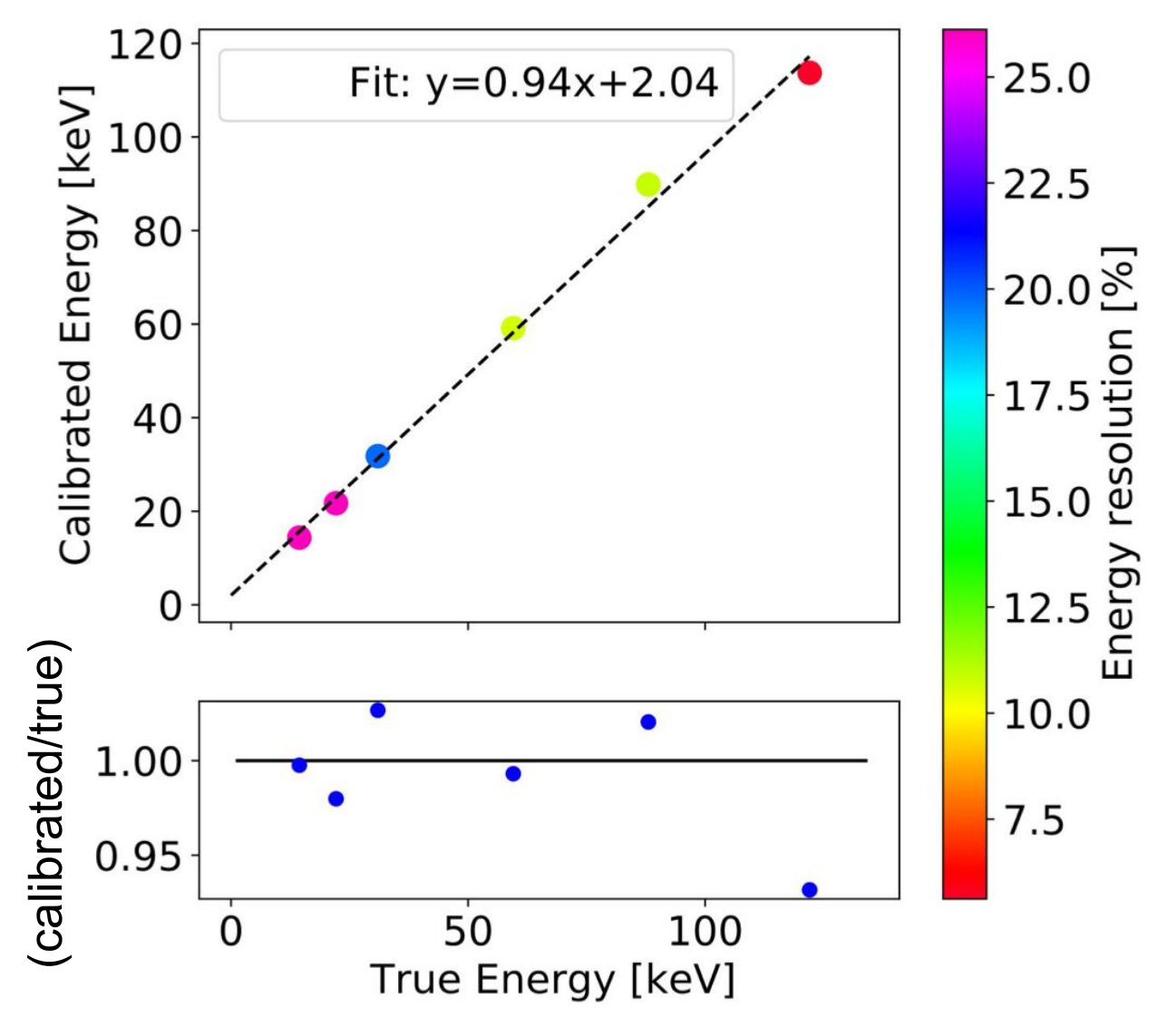}
  \caption{Energy calibration and energy resolution from 14 keV to 122 keV utilizing analog data from one \apone{} pixel \cite{spie}.}
  \label{fig:v1_enres}
\end{subfigure}%
\qquad
\begin{subfigure}{0.47\textwidth}
  \centering
  \includegraphics[width=0.9\linewidth]{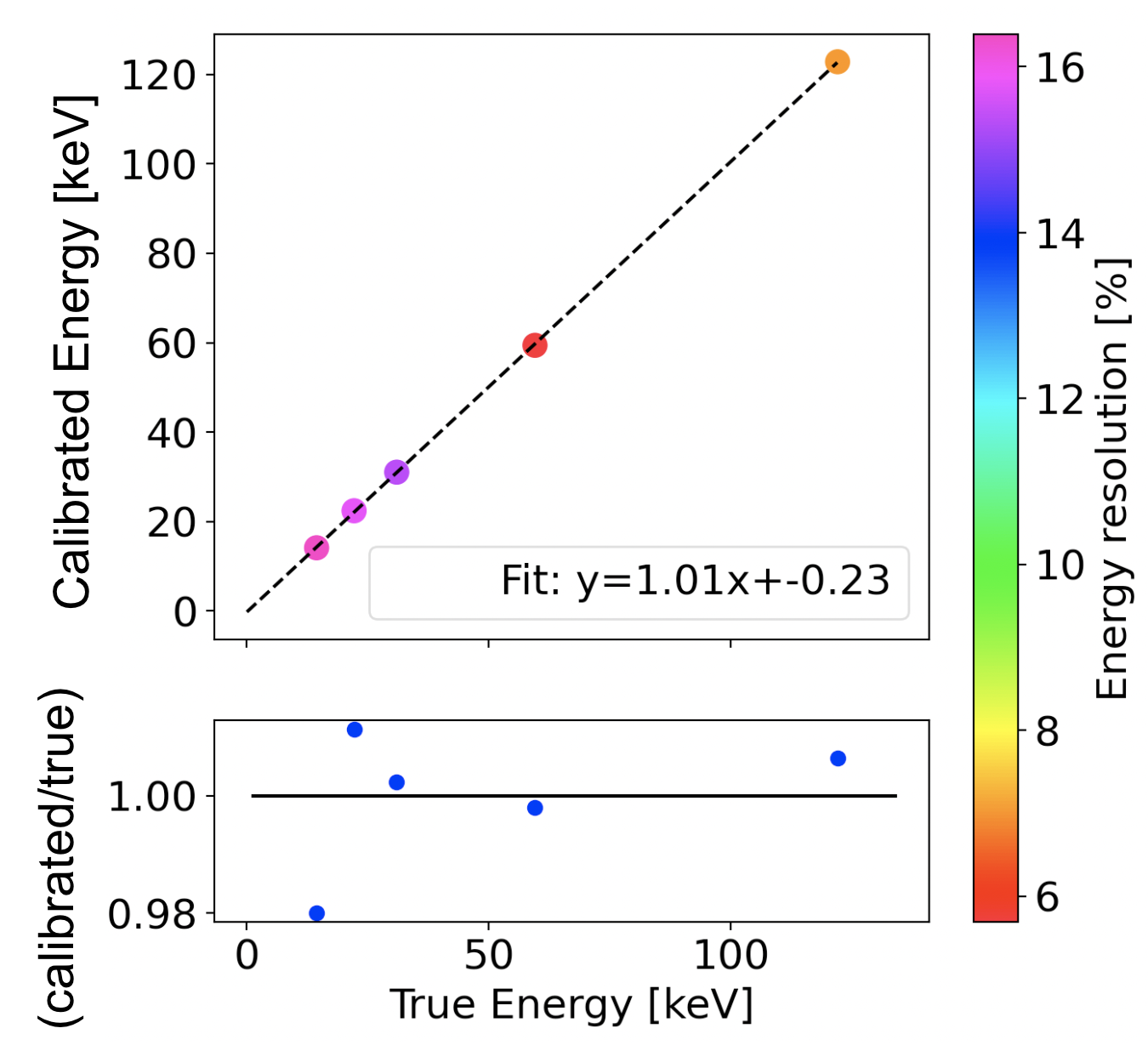}
  \caption{The analog performance of \aptwo{} outperforms that of \apone{} \cite{spie}.}
  \label{fig:v2_enres}
\end{subfigure}%
\caption{Mean and energy resolution (at FWHM) of calibrated spectra for \apone{} and \aptwo{} \cite{spie}.}
\label{fig:enres}
\end{figure}

The space environment and type of incident particle that AstroPix is intended for differs greatly from those in the hadron collider that ATLASPix was designed for, so basic changes to the chip design had to be made. In stepping from ATLASPix to the first $bona-fide$ AstroPix chip, \apone{} or `version1', the digital bit allocation for the time over threshold measurement (Section ~\ref{sec:v2}) was modified so that the precise nanosecond timing resolution was relaxed in favor of energy resolution. The pixels also increased in size. \apone{} was fabricated on 500~\um{} thick silicon wafers with $175\times175$~\um$^2$ in an $18\times18$ array. Insufficient pixel shielding caused oscillations in the digital readout, so only analog data could be read out by \apone. Studies from \cite{spie} detail charge injection studies, threshold studies, and energy calibration performed with \apone. From one probed pixel, energy calibration from analog data was found to match known X-ray and gamma-ray sources within 6\%. A maximum energy resolution (at FWHM, where $E=(2.355\cdot\sigma)/\mu\cdot100\%$ where $\mu$ and $\sigma$ are Gaussian fit parameters from a fit to the photopeak) of 25\% (at 14 keV) was measured (Fig.~\ref{fig:v1_enres}).  

\section{\aptwo{} Testing}
\label{sec:v2}

\begin{figure}
\centering
\begin{subfigure}{0.47\textwidth}
  \centering
  \includegraphics[width=0.9\linewidth]{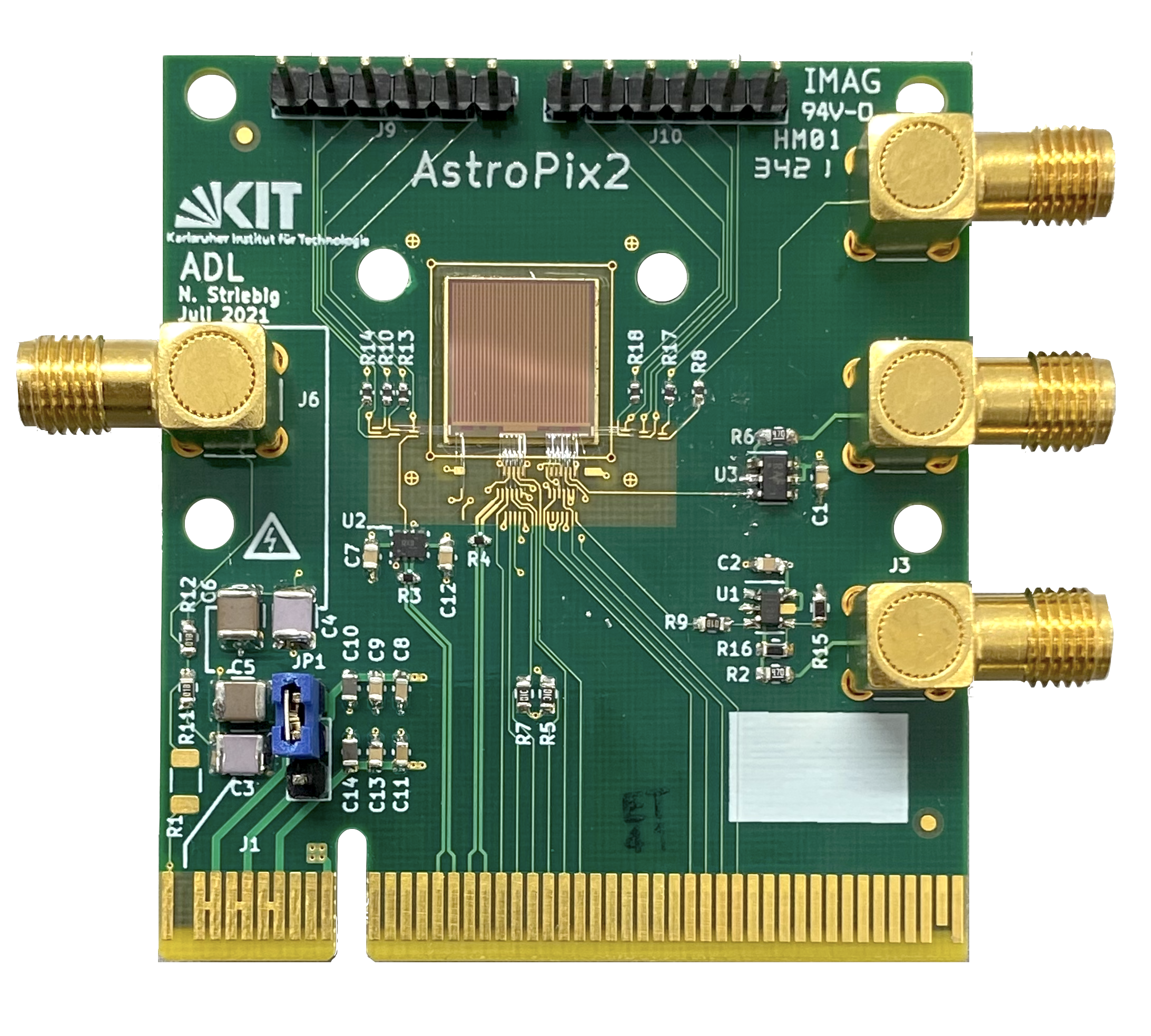}
  \caption{\aptwo{} mounted on a custom printed carrier board.}
  \label{fig:v2board}
\end{subfigure}%
\qquad
\begin{subfigure}{0.47\textwidth}
  \centering
  \includegraphics[width=\linewidth]{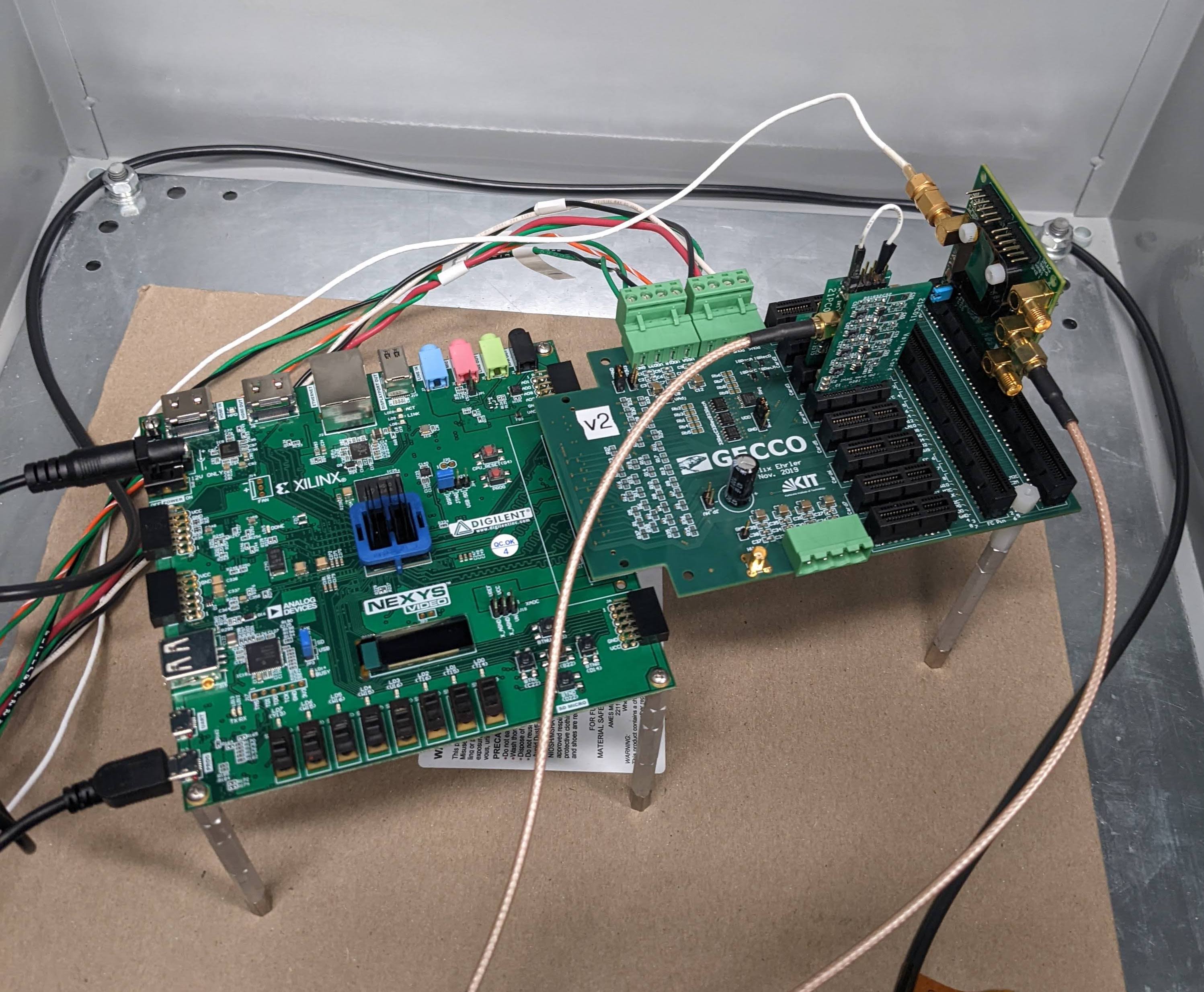}
  \caption{The full AstroPix setup in the lab requires HV supplied to the chip carrier board and measurement made in an aluminum dark box.}
  \label{fig:v2setup}
\end{subfigure}%
\caption{\aptwo{} on the laboratory bench at NASA Goddard Space Flight Center.}
\label{fig:apixSubfig_2}
\end{figure}

\aptwo{} aimed to incrementally move toward a flight prototype by fixing the shielding flaw and redesigning on-pixel circuitry to reduce power consumption. Pixels measure $250\times250$~\um$^2$ in a $35\times35$ array covering an area of $1\times1$~cm$^2$ (Fig.~\ref{fig:v2board}). The guard ring design around each pixel was updated, allowing for higher bias voltages and deeper depletion of the 500~\um{} wafer \cite{spie}. 

The performance of \aptwo{} was studied analogously to \apone, where charge injection studies, threshold studies, and energy calibration was measured from the analog outputs of individual pixels (Fig.~\ref{fig:v2setup}). Energy calibration for \aptwo{} (Fig.~\ref{fig:v2_enres}) is more precise than that of \apone{}, where calibrated photopeak values matched the expected value within $3\%$. \aptwo{} also measures better energy resolution for each calibrated point with a maximum energy resolution at 14 keV of 16\% (25\%) for \aptwo{}~(\apone).

In order to test radiation hardness and performance in a relevant environment, \aptwo{} saw four test-beam campaigns - two at the Fermilab Test Beam Facility \cite{ftbf} with a 120 GeV proton beam and two at the Berkeley 88-Inch Cyclotron \cite{lbnl} with a cocktail of ions up to a linear energy transfer value of 65 MeV cm$^2$/mg. Measurements made during both campaigns confirmed that no catastrophic latchup occurred during running - a state of inactivity where an incident particle triggers a parasitic thyristor, resulting in a short circuit, which can result in runaway current draws and the subsequent destruction of the device. Closer analysis is still underway from the radiation testing with ion beams to determine whether \aptwo{} experienced single event functional interrupt events, where single bits would have been flipped or corrupted by an ion interaction \cite{spie}. The testing in this extreme flux environment also resulted in improved data collection software.

Further details of the design changes from \apone{} to \aptwo, analog performance and tests, and radiation testing can be found in \cite{spie}. 

A major correction from \apone{} to \aptwo{} was the ability to access digital data. While analog data can only be collected from a handful of select pixels, digital data is available from the whole array and is digitized on-chip for readout. In order to save on power and bandwidth, \aptwo{} reads out only row and column information rather than individual pixels, where individual pixel outputs are OR'ed together. In this way, only two channels (row and column) are sent from the array to the digital periphery. The digitized data is returned as an encoded bit stream containing information regarding the time of each hit, whether it is a row or column hit, the location of the row or column, and a Time over Threshold (ToT) measurement. Rather than associating the height of a voltage pulse with the deposited charge, as is done with analog analysis, the digitization requires the input of some voltage threshold and correlates the deposited charge to the amount of time that the resulting voltage pulse was over the threshold. A larger ToT is associated with a larger charge deposit. 

\begin{figure}
    \centering
    \includegraphics[width=0.6\textwidth]{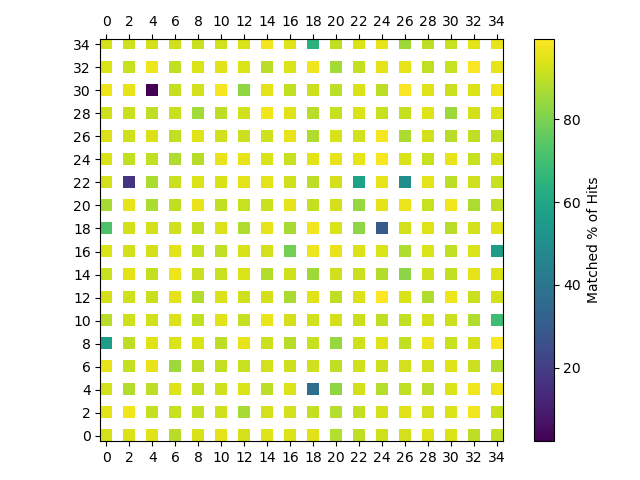}
    \caption{Most of the \aptwo{} array measures both row and column digital hits (as expected) when a signal is injected into the pixel.}
    \label{fig:v2_rowColMatch}
\end{figure}

In this way, the activation of one pixel should return two digitized hits - one for the measurement of charge in the row, and a separate hit for the column. If  corresponding to the same event, these hits must also match in time and measured ToT. This level of correlation was tested with an injected charge administered individually to a sample of pixels around the full array (Fig.~\ref{fig:v2_rowColMatch}). Each pixel was probed 140 times with an injected charge and the fraction of events where two paired row and column hits were recorded is plotted. 99\% of probed pixels read out data where more than 80\% of injections contain matching row and column hits, showing a very high correlation between row and column hits as expected. There is also no noted trend around the array of non-correlating pixels, indicating no large issues with chip fabrication or bonding to its custom printed circuit board. Pixels with poor coincidence may have large rates of noise, confusing the coincidence matching or overwhelming the trigger. Pixels such at these will be masked for data collection campaigns. Future iterations of AstroPix (Section~\ref{sec:future}) will record every pixel individually without OR'ing rows and columns, thus simplifying this problem and eliminating the need for postprocessing coincidence matching. 

\begin{figure}
\centering
\begin{subfigure}{0.47\textwidth}
  \centering
  \includegraphics[width=\linewidth]{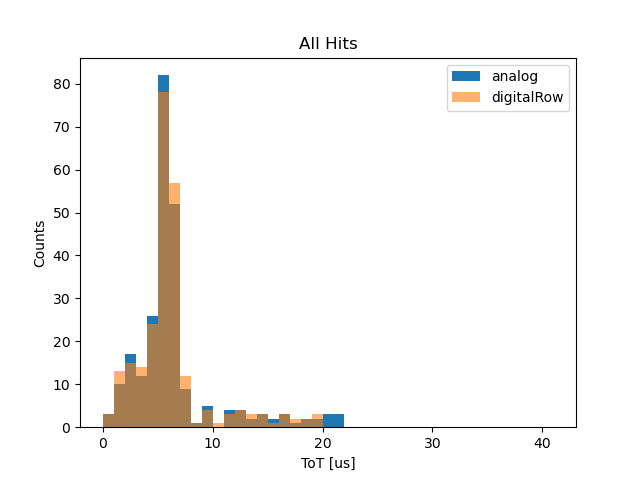}
  \caption{Histograms of analog and digital ToT values from events where a corresponding analog and digital trigger was recorded.}
  \label{fig:da_tot}
\end{subfigure}%
\qquad
\begin{subfigure}{0.47\textwidth}
  \centering
  \includegraphics[width=\linewidth]{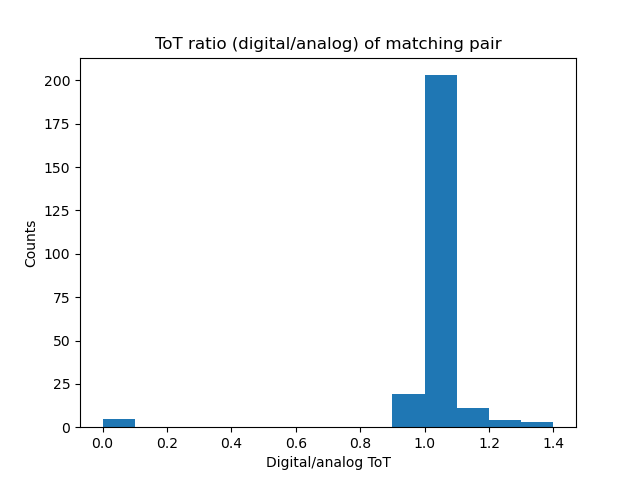}
  \caption{Ratio of digital ToT value to analog ToT-proxy value for events where a corresponding analog and digital trigger was recorded.}
  \label{fig:da_totRatio}
\end{subfigure}%
\caption{Measurements of Barium-133 with a single \aptwo{} pixel confirms that analog and digital data measure identical ToT values.}
\label{fig:v2_digitalAnalog}
\end{figure}

One way to verify the digital circuitry and configuration settings is to measure the analog signal at the same time. The analog signal is read after passing through a lowpass filter while the full digital signal is also passed through a highpass filter, but an incident particle should generate both an analog and a digital signal with nearly identical ToT measurements (or a ToT-proxy measurement as calculated from the shape of the analog voltage pulse). The corner pixel, row 0 column 0, alone was activated on the array and exposed to a 0.01 $\mu$Ci Barium-133 source for 30 minutes. Post-processing software was designed to consider both the output analog and digital data sets, coincide the data in time in order to identify events that triggered both analog and digital readout, and finally to plot this coincident data. Though more analog hits were recorded than digital hits (with a data collection rate of 0.217~Hz compared to 0.162~Hz), 90.6\% of digital data had a corresponding analog hit within a timing window of 0.07~s. The digital ToT measurements of these hits is matched very closely to the corresponding analog ToT-proxy measurement (Fig.~\ref{fig:v2_digitalAnalog}), indicating that chip configuration settings are properly optimized. The timing window of 0.07s was derived as an optimal value from the data and reflects the large latency associated with the analog data collection method - it is not indicative of the inherent timing resolution of the chip. 

This digital output was also used to make a first measurement of depletion depth with \aptwo. The AstroPix design utilizes a thick 500~\um{} wafer and HV bias voltage into order to facilitate a large dynamic range of 25-700 keV. In order to achieve this full depletion, high resistivity wafers of 5~k\ohmcm{} will be utilized but the chips currently under test utilize 300~\ohmcm{} silicon so a smaller depletion depth and therefore dynamic range is expeted.

In order to measure the depletion depth of this lower resistivty array, individual pixels were probed with an Americium-241 source (photopeak at 59.5 keV) and -160V bias voltage. The source is assumed to be point-like, and it is assumed that there is no absorption. A depletion depth $d$ is calculated from the detection rate, 
$$
    r_d = Ap\omega\left(1-e^{-\rho_N\sigma d}\right)~,
$$
where $A$ is the nuclear decay rate (1 MBq), $p$ is the emission probability of 59.5 keV, $\rho_N$ is the number density of silicon, and $\sigma$ is the photoelectric cross section of 59.5 keV in silicon. The geometric factor $\omega$ relates to the pixel size and source distance from the array. The detection rate $r_d$ is found by integrating the measured spectrum of the Americium source.

\begin{figure}
    \centering
    \includegraphics[width=0.5\textwidth]{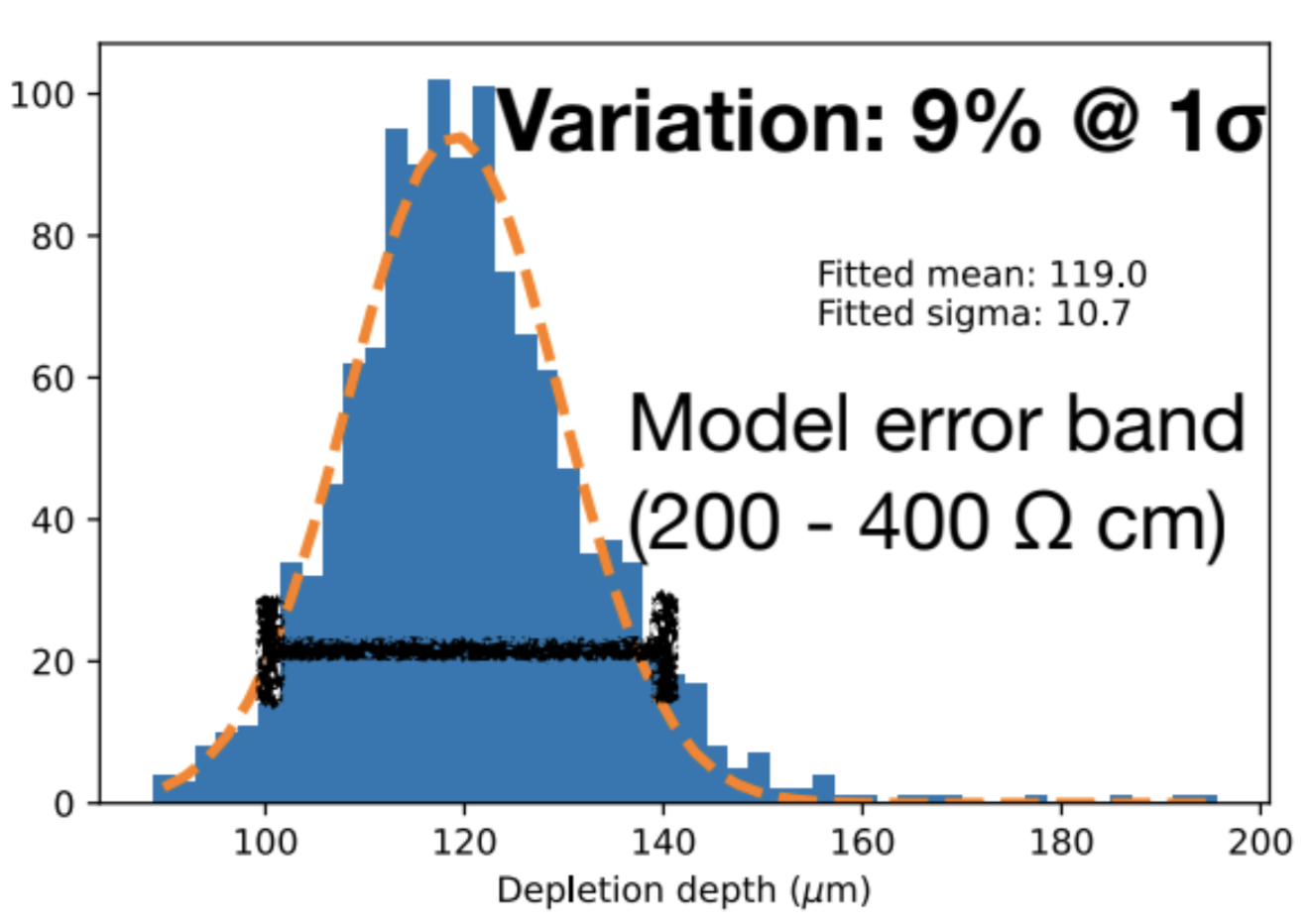}
    \caption{Histogram of depletion depth as measured individually in each \aptwo{} pixel. The dotted orange band is a Gaussian fit with the indicated parameters, and the black bar indicates the expected error on the mean value from the depletion depth model assuming an n-substrate sensor ($\mu$=electron mobility) with resistivity between 200-400~\ohmcm.}
    \label{fig:v2_deplet}
\end{figure}

These direct measurements from every individual pixel of this lower resistivity array show that the depletion achieved with a -160V bias voltage is, on average, 119~\um{} with a 9\% variation at $1\sigma$ (Fig.~\ref{fig:v2_deplet}). The -160V bias value was chosen due to the event rate maximizing and saturating at this bias. Final AstroPix designs will fully deplete 500~\um{} in high-resistivity wafers, and this measured level of depletion on low-resistivity chips offers a promising start.

This value, and the depletion curve over a range of bias values, follows the shape of the model of a p-substrate sensor where 
$$
    d = \sqrt{2\epsilon\mu\rho(V_{bias}+V_i)}~,
$$
where $\epsilon$ is the permittivity $1.04\times10^{-12}$~F/cm, $\mu$ is hole mobility (500~cm$^2$/s/V), $\rho$ is the sensor resistivity of 300~\ohmcm, and $V_i$ is a built-in potential of 0.6~V. At the time of writing, a systematic offset is found between the model prediction and data such that the data agrees more closely with the modeling of an n-substrate sensor where $\mu$ is the electron mobility of 1500~cm$^2$/s/V. Work is underway to further understand this effect.

\section{Ongoing work and Next Steps}
\label{sec:future}

Testing of \aptwo{} is still underway, with a current emphasis on digital data read out from the full array. There is also active testing of \aptwo{} fabricated on silicon wafers with high resistivity. This higher resistivity of 5~k\ohmcm{} as compared to the 300~\ohmcm{} wafers utilized for the data thus far presented should allow for larger depletion depth and therefore enhanced charge collection and dynamic range. \aptwo{} will see another Fermilab Test Beam Facility campaign in early 2023 with emphasis on multiple-layer readout utilizing multiple \aptwo{} chips in the beamline and testing preliminary particle tracking software.

\begin{figure}
\centering
\begin{subfigure}{0.47\textwidth}
  \centering
  \includegraphics[width=0.8\linewidth]{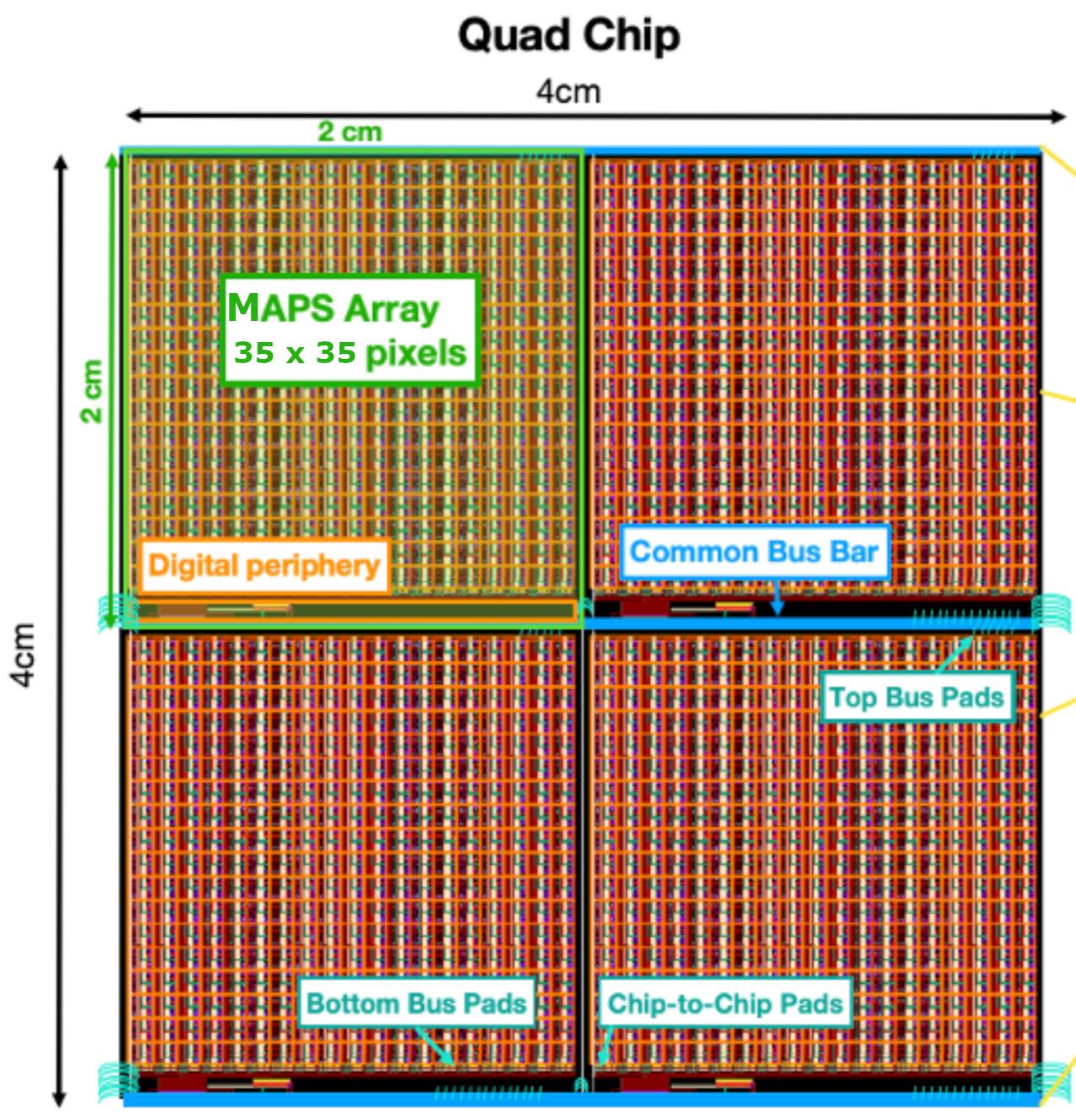}
  \caption{\apthree{} (fabricated January 2023) will be diced as a quad chip with four AstroPix arrays and common readout.}
  \label{fig:quad}
\end{subfigure}%
\qquad
\begin{subfigure}{0.47\textwidth}
  \centering
  \includegraphics[width=0.8\linewidth]{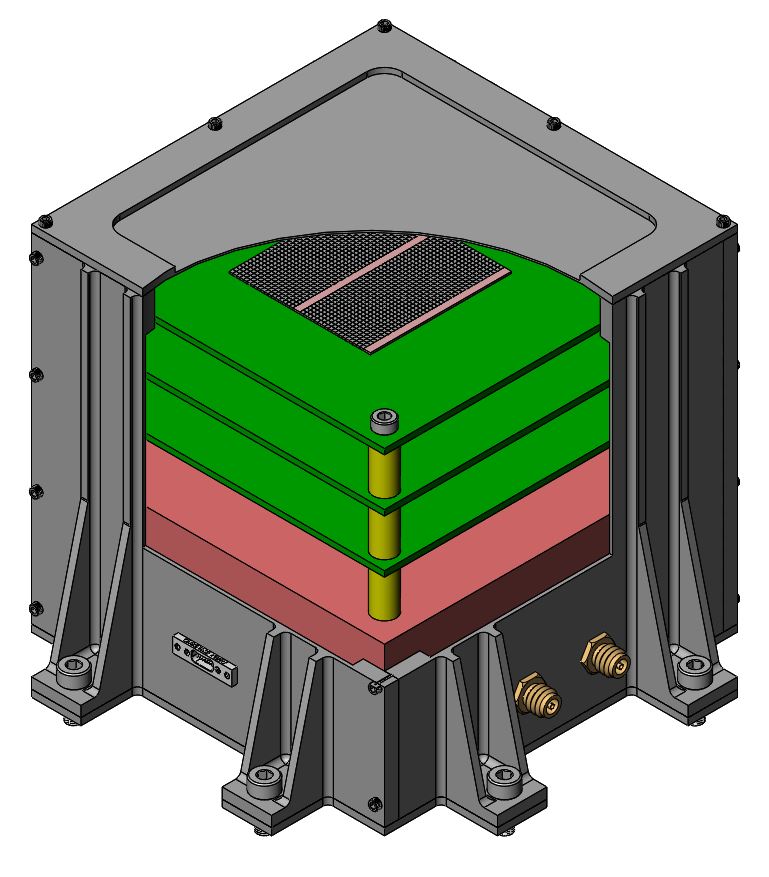}
  \caption{A-STEP is a technology demonstration payload with three layers of \apthree{} quad chips intended to fly on a sounding rocket in late 2024. The preliminary 1U design is shown but the final design will likely be 2U to accomodate all flight components.}
  \label{fig:astep}
\end{subfigure}%
\caption{\apthree{} is currently being fabricated and will be diced as a quad-chip. This version will be utilized in a sounding rocket payload, A-STEP.}
\label{fig:v3}
\end{figure}

Concurrently, the AstroPix project continues to look ahead with the design of \apthree. This is a flight-protype version with a $35\times35$ array of $500\times500$~\um$^2$ pixels with a $300\times300$~\um$^2$ active area. The continued increase in pixel size does not impact the energy resolution of future missions and allows AstroPix to consume less power and bandwidth with a smaller number of channels overall. However, the larger pixels create engineering challenges - for example, the large active area creates high capacitance levels which increase noise. The \apthree{} strategy of reducing the active pixel area with respect to the pixel pitch avoids these complications. \apthree{} will also be diced from the wafer as a $4\times4$~cm$^2$ quad chip, where four \apthree{} arrays will be connected through common bus bars (Fig.~\ref{fig:quad}). \apthree{} was delivered from the foundry in January 2023 and testing began shortly thereafter. 

The \apthree{} quad chip will be used on the Astropix Sounding rocket Technology dEmonstration Payload (A-STEP), currently scheduled for launch in late 2024. The 2U payload will consist of three layers of \apthree{} quad chips with a thin aluminum housing (Fig.~\ref{fig:astep}), along with supporting electronics. A sounding rocket flight of roughly 10 minutes will take A-STEP 500~km above ground and provide the opportunity to measure cosmic rays and gamma rays. The project intention is to demonstrate functionality of the AstroPix sensors in a relevant space environment by reconstructing these charged particle tracks. The A-STEP project kicked off in October 2022 led through Goddard Space Flight Center with support from Wallops Flight Facility engineering support and the Sounding Rocket Program Office for coordination and planning.

\begin{figure}
\centering
\begin{subfigure}{0.47\textwidth}
  \centering
  \includegraphics[width=0.9\linewidth]{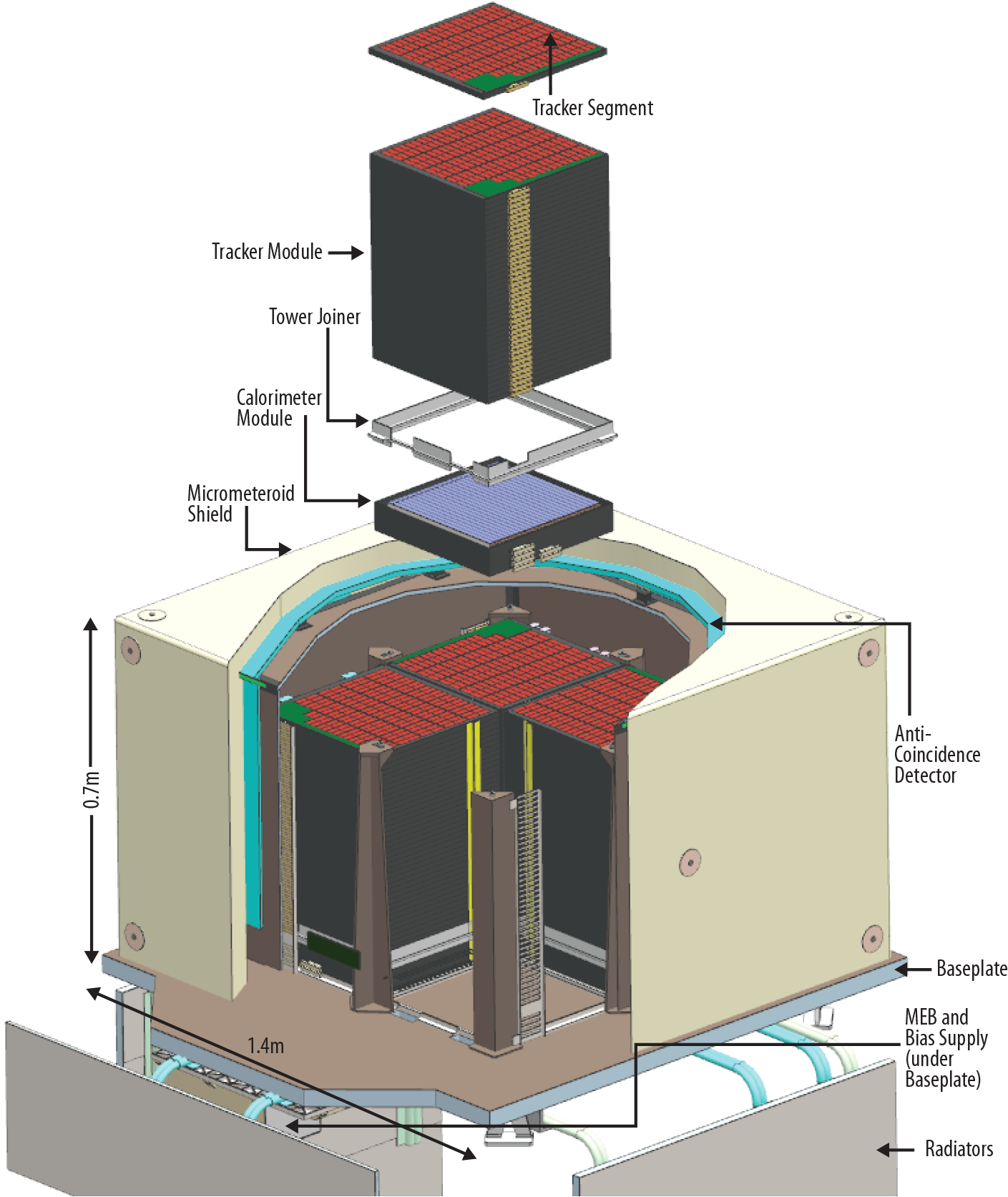}
  \caption{Exploded view of \amx{}, utilizing an AstroPix-based tracker.}
  \label{fig:amx}
\end{subfigure}%
\qquad
\begin{subfigure}{0.47\textwidth}
  \centering
  \includegraphics[width=\linewidth]{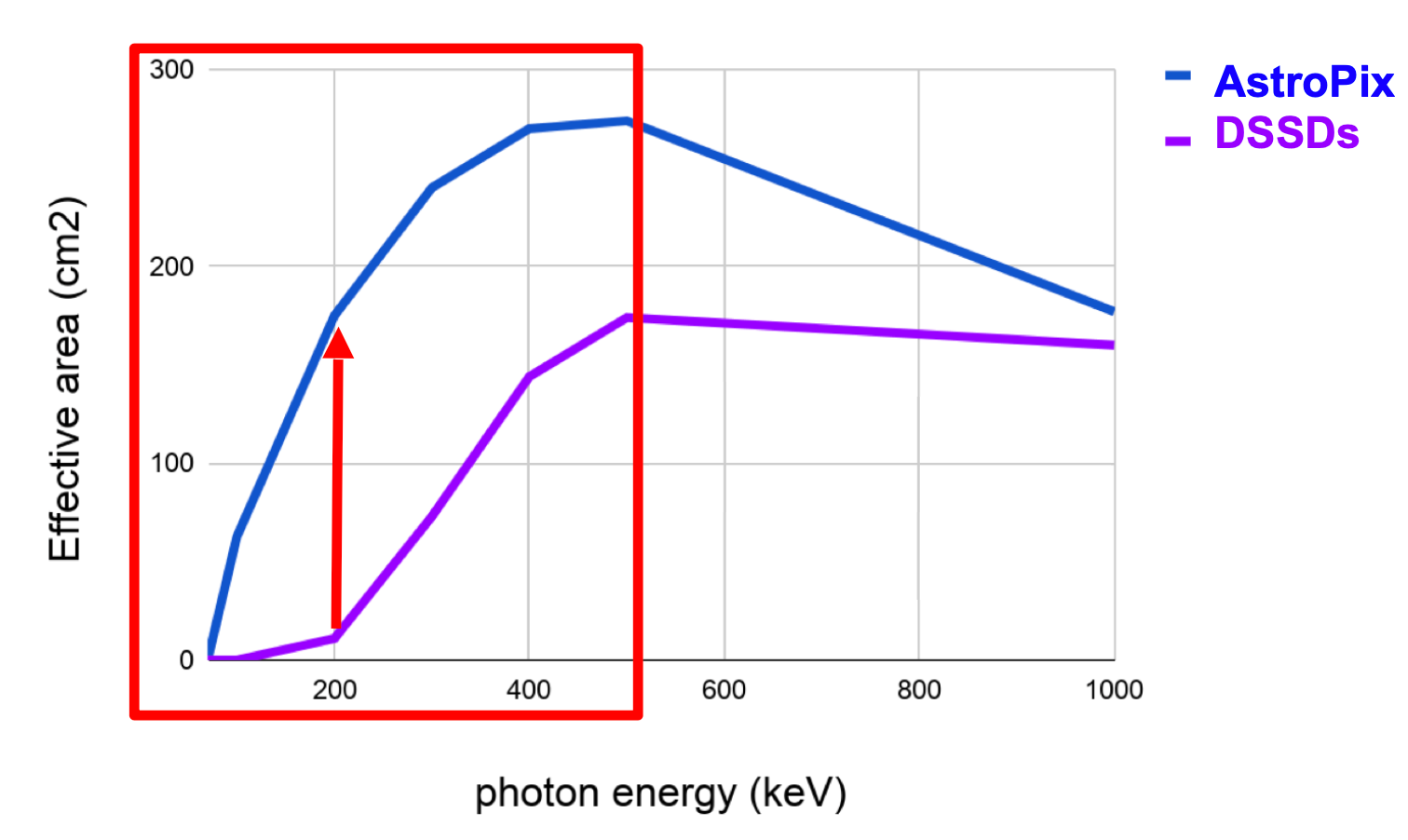}
  \caption{The use of AstroPix provides \amx{} a factor of 10 improvement in effective area over double sided silicon strip detectors from 50-500 keV.}
  \label{fig:dssds}
\end{subfigure}%
\caption{Future implementation of AstroPix in the MIDEX-scale \amx{} tracker will increase effective area at low energies \cite{amx_paper}.}
\label{fig:future}
\end{figure}

AstroPix is a mission enabling technology for future large-scale gamma-ray instruments. It is implemented in the 2021 MIDEX concept \amx{} as the main detector of the tracking subsystem \cite{amx_paper}. The \amx{} tracker design features four identical towers of 40 layers of AstroPix with 95 quad chips per layer (Fig.~\ref{fig:amx}). The use of AstroPix at this scale provides \amx{} a significant improvement of effective area at low energies (50-500 keV) over that with the use of double sided silicon strip detectors (Fig.~\ref{fig:dssds}). 

\begin{figure}
    \centering
    \includegraphics[width=0.4\textwidth]{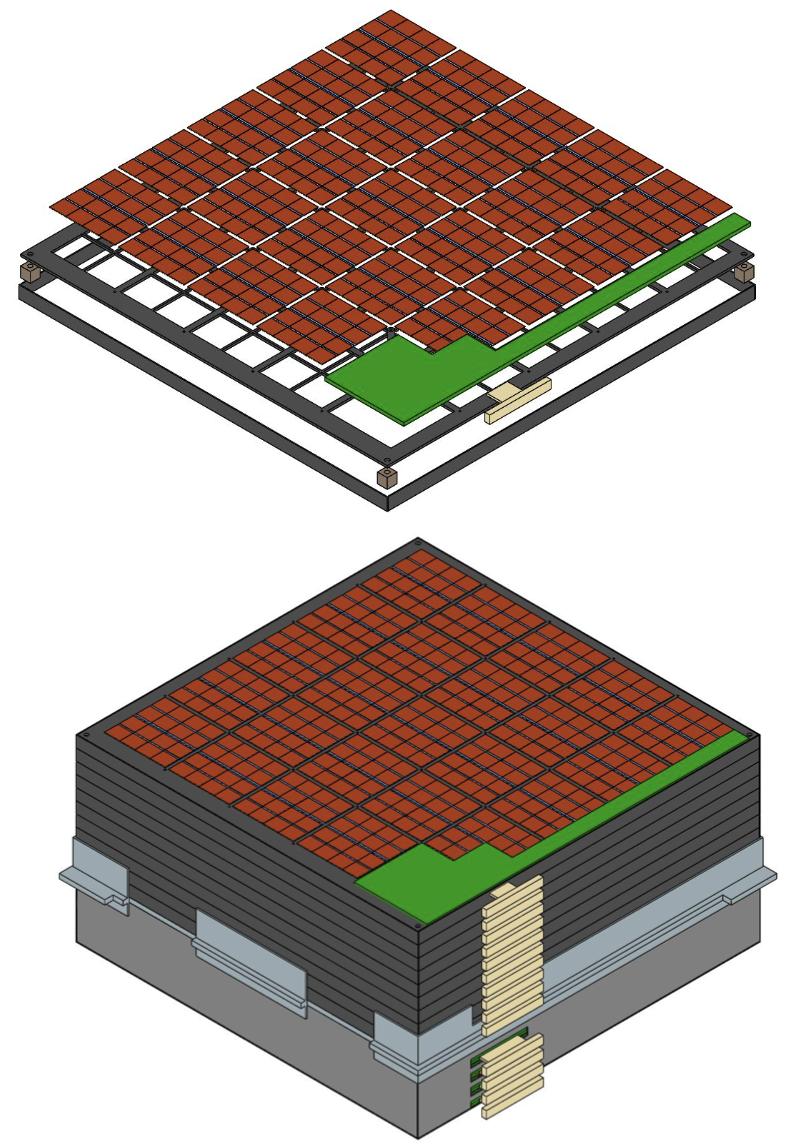}
    \caption{An \amx{} prototype will be built as a balloon instrument and utilize AstroPix as the tracker.}
    \label{fig:compair}
\end{figure}

In order to build to full implementation in \amx, an \amx{} prototype is planned. This prototype is a combination of many efforts, including: AstroPix (as a novel tracker), a cesium-iodide calorimeter, the ComPair balloon instrument \cite{compair}, and Compton event reconstruction improvements. The \amx{} prototype (Fig.~\ref{fig:compair}), a next-generation ComPair instrument, will be one tower of the \amx{} instrument (though with less AstroPix tracker layers). The aim is to demonstrate operation of the full system in a relevant environment.  

These implementations will also utilize future versions of AstroPix. AstroPix\_v4 is currently in development with planned upgrades to the ToT readout system, and individual pixel readout (without row and column OR'ing) and threshold tuning.

\section{Summary and Outlook}
\label{sec:summary}

Future large-scale gamma-ray instruments, as prioritized by the Astro2020 Decadal Survey, will benefit from new technologies that allow for measurements with low noise and precise position and energy resolution. As an HVCMOS sensor, AstroPix serves as mission-enabling technology for these next-generation instruments. With development rooted in work done by the collider-based particle physics community, AstroPix has now realized two design iterations with a third recently delivered at the time of writing. The sensors are capable of analog and digital data readout and measure energy resolutions from analog data better than 16\% at 14 keV in the most recent design iteration, \aptwo. Digital readout is being tested with \aptwo, where individual pixels are returning triggered data uniformly around the array as expected. The next design iteration, \apthree, will be diced as a quad chip sharing common readout of four individual arrays connected via a common bus bar and be utilized in future technology demonstrations including the sounding rocket payload A-STEP and \amx{} prototype ComPair balloon instrument. 

The international AstroPix team spans three countries over five institutions with more than half of the contributors being early career scientists or engineers (including students). The team looks forward to continued testing of AstroPix, future development, and further implementation in next-generation space-based instruments.   

\newpage
\bibliographystyle{plain}
\bibliography{astropix} 

\end{document}